\documentclass[11pt]{revtex4}
\usepackage{amssymb,epsf}
\usepackage{latexsym}

\begin{document}

\title{Asymptotic charged BTZ black hole solutions}
\author{S. H. Hendi\footnote{email address: hendi@shirazu.ac.ir}}
\affiliation{Physics Department and Biruni Observatory, Shiraz University, Shiraz 71454,
Iran\\
Research Institute for Astrophysics and Astronomy of Maragha
(RIAAM), Maragha, Iran, P.O. Box 55134-441}

\begin{abstract}
The well-known $(2+1)$-dimensional Reissner-Nordstr\"{o}m (BTZ) black hole
can be generalized to three dimensional Einstein-nonlinear electromagnetic
field, motivated from obtaining a finite value for the self-energy of a
pointlike charge. Considering three types of nonlinear electromagnetic
fields coupled with Einstein gravity, we derive three kinds of black hole
solutions which their asymptotic properties are the same as charged BTZ
solution. In addition, we calculate conserved and thermodynamic quantities
of the solutions and show that they satisfy the first law of thermodynamics.
Finally, we perform a stability analysis in the canonical ensemble and show
that the black holes are stable in the whole phase space.
\end{abstract}

\maketitle

%%%%%%%%%%%%%%%%%%%%%%%%%%%%%%%%%%%%%%%%%%%%%%%%%%%%%%%%%%%%%%%%

\section{Introduction}

Nonlinear field theories are of interest to different branches of
mathematical physics because most physical systems are inherently nonlinear
in nature. Nonlinear action of Abelian gauge theories have been considered
in the context of superstring theory. In fact, it has shown that \cite%
{Fradkin85} all order loop corrections to gravity may be considered as a
Born-Infeld (BI) type Lagrangian \cite{BI}. Also, the dynamics of D-branes
and some soliton solutions of supergravity are governed by the BI action
\cite{Leigh89}. The first attempt to relate the nonlinear electrodynamics
(NLEDs) and gravity was made by Hoffmann \cite{Hoffmann}. Considering NLEDs
coupled to the gravitational field (with or without scalar field) may lead
to black hole solutions with interesting properties \cite%
{BIpaper,PMIpaper,Ayon98,Oliveira94,Soleng}.

In addition to the nonlinear BI, other types of NLEDs have been studied in a
number of papers \cite{PMIpaper,Soleng}. It is known that the NLEDs proposed
by Born and Infeld had the aim of obtaining a finite value for the
self-energy of a point-like charge. So in this paper, we consider three
kinds of BI-type fields with the following motivations:

First, modifying the linear Maxwell theory to NLED theory may eliminate the
problem of divergency in electromagnetic field.

Second, it is notable that one can find regular black hole solutions of the
Einstein field equations coupled to a suitable NLEDs \cite%
{Oliv94,Soleng,Palatnik98,Ayon46,Ayon84,Ayon99}. Also, an interesting
property which is common to all the NLED models is the fact that these
models satisfy the zeroth and first laws of black hole mechanics \cite%
{Rasheed97}.

Third, it is also remarkable that, BI-type theories are singled
out among the classes of NLEDs by their special properties such as
the absence of shock waves, birefringence phenomena \cite{Boillat}
and enjoying an electric-magnetic duality \cite{GibRash}.

Fourth, the appropriate world-volume dynamics on a curved $D3$-brane may
provide a plausible frame-work at Planck scale by incorporating the
Einstein-NLEDs. At this point of time, elimination of strong intrinsic
curvature in the regime by the strong nonlinearity in the $U(1)$ gauge
theory is remarkable \cite{Gopakumar,Ayon99,Tamaki2000,Kar05}.

Fifth, from the point of view of AdS/CFT correspondence in hydrodynamic
models, it has been shown that, unlike gravitational correction,
higher-derivative terms for Abelian fields in the form of NLED do not affect
this ratio \cite{Brigante,Kovtun,CaiJHEP08,GeJHEP08}. In addition, in
applications of the AdS/CFT correspondence to superconductivity, NLED
theories make crucial effects on the condensation \cite{Jing10} as well as
the critical temperature of the superconductor and its energy gap \cite%
{Gregory09,Pan10}.

Sixth, it has been shown that \cite{Salim55,Salim25} the effects of NLED
become important when we investigate super-strongly magnetized compact
objects, such as pulsars, neutron stars, magnetars and strange quark
magnetars. In addition, it has proved that \cite{Salim55} if one consider a
NLED to incorporate into the photon dynamics, Gravitational Redshift depends
on the magnetic field pervading the pulsar (while the Gravitational Redshift
is independent of any background magnetic field in general relativity).
Also, since the Gravitational Redshift of magnetized compact objects is
connected to the mass--radius relation of the objects, it is important to
note that NLED affects on the mass--radius relation of the objects.

Motivated by the recent results mentioned above, we study black hole
solutions in Einstein gravity with negative cosmological constant coupled to
NLED theory. Considering these nonlinear fields in $(2+1)$-dimensional
spacetime, with asymptotic BTZ (Banados--Teitelboim--Zanelli) behavior, help
us to find a simple mechanism for analyzing the properties of the solutions.

Physicists believe that one of the great achievements in gravity is
discovery of the $(2+1)$-dimensional BTZ black hole solutions \cite%
{BTZ1,BTZ2,BTZ3}. In fact, BTZ black holes provide a simplified model to
investigate and find some conceptual issues such as black hole
thermodynamics \cite{Carlip95,Ashtekar02,Sarkar06}, quantum gravity, string
and gauge theory and specially, in the context of the AdS/CFT conjecture
\cite{Witten98,Carlip05}. Furthermore, BTZ solutions perform a central role
to improve our perception of gravitational interaction in low dimensional
spacetime \cite{Witten07}. Generalization of BTZ black hole and its
properties to higher dimensions and also its near-horizon solutions have
been studied in \cite%
{Thermodynamics,Saida99,Cadoni08,Larranaga10,Hyun97,Sfetsos98,Canfora10,Claessens09,BTZlike}%
.

Since in this paper, we consider three classes of the Einstein-NLED field
and expect to obtain asymptotic BTZ black hole, we want to discuss about two
significant properties of charged BTZ solutions. First, it is notable that
for $(2+1)$-dimensional charged BTZ solutions, $A_{t}$ and the
electromagnetic field are proportional to $\ln r$ and $r^{-1}$,
respectively. Second, the charge term of the laps function in horizon flat $%
(2+1)$-dimensional BTZ solutions, is a logarithmic function of $r$.

Organization of the paper is as follows: at first, we give a brief review of
the field equations of Einstein gravity sourced by the NLED field. Then we
consider three dimensional spacetime and find relative solutions. After that
we investigate their properties, especially singularity and asymptotic
behavior of them. Then, we obtain conserved and thermodynamic quantities of
the black holes, in which satisfy the first law of thermodynamics. Also, we
analyze the local stability in canonical ensemble and at last we confirm
that obtained solutions are asymptotic BTZ. We finish our paper with some
conclusions.

\section{$(2+1)$-dimensional black holes with NLED field}

The $(2+1)$-dimensional action of Einstein gravity with NLED field in the
presence of cosmological constant is given by
\begin{equation}
I_{G}=-\frac{1}{16\pi }\int_{\mathcal{M}}d^{3}x\sqrt{-g}\left[ R-2\Lambda +L(%
\mathcal{F})\right] -\frac{1}{8\pi }\int_{\partial \mathcal{M}}d^{2}x\sqrt{%
-\gamma }K,  \label{Act}
\end{equation}
where ${R}$ is the Ricci scalar, $\Lambda $ refers to the negative
cosmological constant which in general is equal to $-1/l^{2}$ for
asymptotically anti-deSitter solutions, in which $l$ is a scale length
factor. In Eq. (\ref{Act}), $L(\mathcal{F})$ is the Lagrangian of NLED
field. Here, we consider three classes of Born-Infeld-like NLED fields,
namely Born-Infeld nonlinear electromagnetic (BINEF), Exponential form of
nonlinear electromagnetic field (ENEF) and Logarithmic form of nonlinear
electromagnetic field (LNEF) in which their Lagrangians are
\begin{equation}
L(\mathcal{F})=\left\{
\begin{array}{ll}
4\beta ^{2}\left( 1-\sqrt{1+\frac{\mathcal{F}}{2\beta ^{2}}}\right) , &
\text{BINEF} \\
\beta ^{2}\left( \exp (-\frac{\mathcal{F}}{\beta ^{2}})-1\right) , & \text{%
ENEF} \\
-8\beta ^{2}\ln \left( 1+\frac{\mathcal{F}}{8\beta ^{2}}\right) , & \text{%
LNEF}%
\end{array}
\right. .  \label{LagEM}
\end{equation}
In this equation, $\beta $ is called the nonlinearity parameter, the Maxwell
invariant $\mathcal{F}=F_{\mu \nu }F^{\mu \nu }$ in which $F_{\mu \nu
}=\partial _{\mu }A_{\nu }-\partial _{\nu }A_{\mu }$ is the electromagnetic
field tensor and $A_{\mu }$ is the gauge potential.

It is natural to expect that the nonlinear electromagnetic field appears as
an effective theory of string/M---theory. For instance, one of the subgroup
of the $E_{8} \times E_{8}$ or $SO(32)$ gauge group is $U(1)$. Ignoring all
other gauge fields leaves us with the effective quartic order of $U(1)$
Lagrangian, $(F_{\mu \nu}F^{\mu \nu})^2$ \cite%
{Liu26,Gross87,Bergshoeff,Chemissany07}. In addition, Natsuume \cite%
{Natsuume} considered the next order correction terms in the heterotic
string effective action of a magnetically-charged black hole \cite{GHS} and
obtained the $(F_{\mu \nu}F^{\mu \nu})^2$ term as a subset of all possible $%
\alpha$ corrections to the bosonic sector of supergravity, which is the same
order as the Gauss-Bonnet term \cite{Liu26}
\begin{equation}
L_{cor}=\alpha \left[ a\left( R_{\mu \nu \rho \sigma}R^{\mu \nu \rho
\sigma}-4R_{\mu \nu}R^{\mu \nu}+R^2 \right)+b\left(F_{\mu \nu}F^{\mu
\nu}\right)^2 \right].
\end{equation}
Furthermore, calculating one-loop approximation of QED, it has shown that
\cite{Ritz} the effective Lagrangian is given by
\begin{equation}
L_{eff}= c\left( F_{\mu \nu}F^{\mu \nu} \right)+d\left(F_{\mu \nu}F^{\mu
\nu}\right)^2.
\end{equation}

Euler and Heisenberg have shown that the correction contain
logarithmic form in the electromagnetic field strength appear in
the calculation of exact $1$-loop corrections for electrons in a
uniform electromagnetic field background \cite{Heisenberg}, which
is description of vacuum polarization effects. So, these
corrections are a kind of effective actions in quantum
electrodynamics.

Furthermore, logarithmic form of the electrodynamic Lagrangians, like BI
electrodynamics, removes divergences in the electric field. Although the
exponential form of BI-like NLED does not cancel the divergency of electric
field at $r=0$, but its singularity is too much weaker than in, e.g.,
Einstein-Maxwell gravity.

From the cosmological point of view, these BI-like NLEDs have also
been used to explain an equation of state of radiation for
inflation \cite{Altshuler}. As an example of a BI-like Lagrangian
with a logarithmic term, four dimensional asymptotically flat
solutions of Einstein gravity was discussed in \cite{Soleng}.
Expanding $L(\mathcal{F}) $'s for large values of $\beta$, one can
write
\begin{equation}
\left. L(\mathcal{F})\right\vert _{\text{Large }\beta }=\left\{
\begin{array}{ll}
-\mathcal{F}+\frac{\mathcal{F}^{2}}{8\beta ^{2}}-\frac{\mathcal{F}^{3}}{%
32\beta ^{4}}+\frac{5\mathcal{F}^{4}}{512\beta ^{6}}+O\left( \frac{1}{\beta
^{8}}\right) , & \text{BINEF} \\
-\mathcal{F}+\frac{\mathcal{F}^{2}}{2\beta ^{2}}-\frac{\mathcal{F}^{3}}{%
6\beta ^{4}}+\frac{\mathcal{F}^{4}}{24\beta ^{6}}+O\left( \frac{1}{\beta ^{8}%
}\right) , & \text{ENEF} \\
-\mathcal{F}+\frac{\mathcal{F}^{2}}{16\beta ^{2}}-\frac{\mathcal{F}^{3}}{%
192\beta ^{4}}+\frac{\mathcal{F}^{4}}{2048\beta ^{6}}+O\left( \frac{1}{\beta
^{8}}\right) , & \text{LNEF}%
\end{array}
\right.  \label{LagEm2}
\end{equation}
which confirm that $L(\mathcal{F})$'s reduce to the standard Maxwell form $L(%
\mathcal{F})=-\mathcal{F}$, for $\beta \longrightarrow \infty $ and also the
leading first correction of Maxwell theory has $\left(F_{\mu \nu}F^{\mu
\nu}\right)^2$ form.

Furthermore, we should note that the second integral in Eq. (\ref{Act}) is
the Gibbons-Hawking surface term \cite{GibHaw} which is chosen such that the
variational principle is well defined. The factor $K$ is the trace of the
extrinsic curvature $K_{ab}$ of any boundary ${\partial \mathcal{M}}$ of the
manifold ${\mathcal{M}}$, with induced metric $\gamma _{ab}$. Although three
dimensional solution of Einstein-BI gravity has been investigated before
\cite{BIBTZ}, but we present it again with two motivations: firstly, one can
confirm that our BI solution is more compact and we discuss about the
geometry of BI black hole and its horizon precisely, and secondly, in order
to compare the solutions of the exponential and logarithmic Lagrangian with
BI solution, we need to present BI solution here.

Varying the action (\ref{Act}) with respect to the gravitational
field $g_{\mu \nu }$ and the gauge field $A_{\mu }$, the field
equations are obtained as
\begin{equation}
R_{\mu \nu }-\frac{1}{2}g_{\mu \nu }\left( R-2\Lambda \right) =\alpha T_{\mu
\nu },  \label{FE1}
\end{equation}
\begin{equation}
\partial _{\mu }\left( \sqrt{-g}L_{\mathcal{F}}F^{\mu \nu }\right) =0,
\label{FE2}
\end{equation}
where
\begin{equation}
T_{\mu \nu }=\frac{1}{2}g_{\mu \nu }L(\mathcal{F})-2F_{\mu \lambda }F_{\nu
}^{\;\lambda }L_{\mathcal{F}},  \label{FE3}
\end{equation}
and $L_{\mathcal{F}}=\frac{dL(\mathcal{F})}{d\mathcal{F}}$. Our
main aim here is to obtain charged static black hole solutions of
the field equations (\ref{FE1}) - (\ref{FE3}) and investigate
their properties. We assume $(2+1)$-dimensional metric has the
following form
\begin{equation}
ds^{2}=-f(r)dt^{2}+\frac{dr^{2}}{f(r)}+r^{2}d\theta ^{2}  \label{Metric}
\end{equation}

Using the gauge potential ansatz $A_{\mu }=h(r)\delta _{\mu }^{0}$ in the
NLED fields equation (\ref{FE2}) leads to the following differential
equations
\begin{equation}
\begin{array}{rr}
rh^{\prime \prime }(r)+h^{\prime }(r)\left[ 1-\left( \frac{h^{\prime }(r)}{%
\beta }\right) ^{2}\right] =0, & \text{BINEF}\vspace{0.1cm} \\
r\left[ 1+\frac{4}{\beta ^{2}}h^{\prime }(r)\right] h^{\prime \prime
}(r)+h^{\prime }(r)=0, & \text{ENEF}\vspace{0.1cm} \\
r\left[ 1+\left( \frac{h^{\prime }(r)}{2\beta }\right) ^{2}\right] h^{\prime
\prime }(r)+h^{\prime }(r)\left[ 1-\left( \frac{h^{\prime }(r)}{2\beta }%
\right) ^{2}\right] =0, & \text{LNEF}
\end{array}
,  \label{heq}
\end{equation}
with the following solutions
\begin{equation}
h(r)=\frac{q}{2}\times \left\{
\begin{array}{ll}
2\ln \left[ \frac{r}{2l}\left( 1+\Gamma \right) \right] , & \text{BINEF}%
\vspace{0.1cm} \\
\frac{\beta r\sqrt{L_{W}}}{q}+{Ei}\left( 1,\frac{L_{W}}{2}\right) +\ln
\left( \frac{2q^{2}}{l^{2}\beta ^{2}}\right) +\gamma -2, & \text{ENEF}%
\vspace{0.1cm} \\
2\ln \left[ \frac{r}{2l}\left( 1+\Gamma \right) \right] -\frac{2r^{2}\beta
^{2}}{q^{2}}\left( 1-\Gamma \right) -1, & \text{LNEF}%
\end{array}
\right. ,  \label{h(r)}
\end{equation}
where $\Gamma =\sqrt{1+\frac{q^{2}}{r^{2}\beta ^{2}}}$, $q$ is an
integration constant which is related to the electric charge of
the black hole. In addition,
$L_{W}=LambertW(\frac{4q^{2}}{r^{2}\beta ^{2}})$ which satisfies
$LambertW(x)\exp \left[ LambertW(x)\right] =x$, $\gamma =\gamma
(0)\simeq 0.57722$ and the special function ${Ei}\left( 1,x\right)
=\int\limits_{1}^{\infty }\frac{e^{-xz}}{z}dz$ (for more details,
see \cite{Lambert}).

It is easy to show that the non-vanishing components of the electromagnetic
field tensor can be written in the following form
\begin{equation}
F_{tr}=\frac{q}{r}\times \left\{
\begin{array}{ll}
\Gamma ^{-1}, & \text{BINEF} \\
\frac{r\beta \sqrt{L_{W}}}{2q}, & \text{ENEF} \\
\frac{2\beta ^{2}r^{2}}{q^{2}}\left( \Gamma -1\right) , & \text{LNEF}%
\end{array}
\right. .  \label{Ftr}
\end{equation}

\begin{figure}[tbp]
$
\begin{array}{ccc}
\epsfxsize=5cm \epsffile{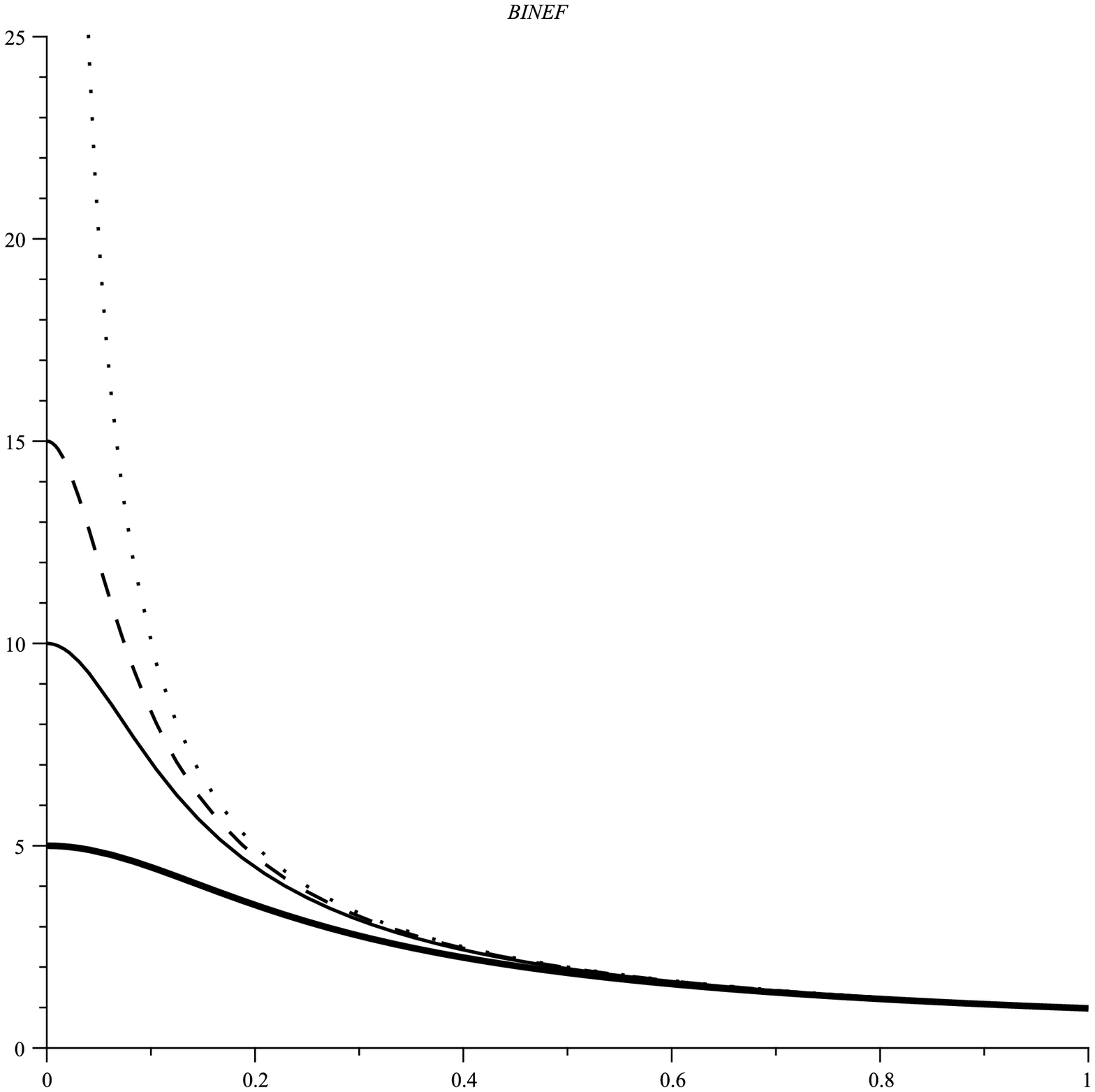} & \epsfxsize=5cm
\epsffile{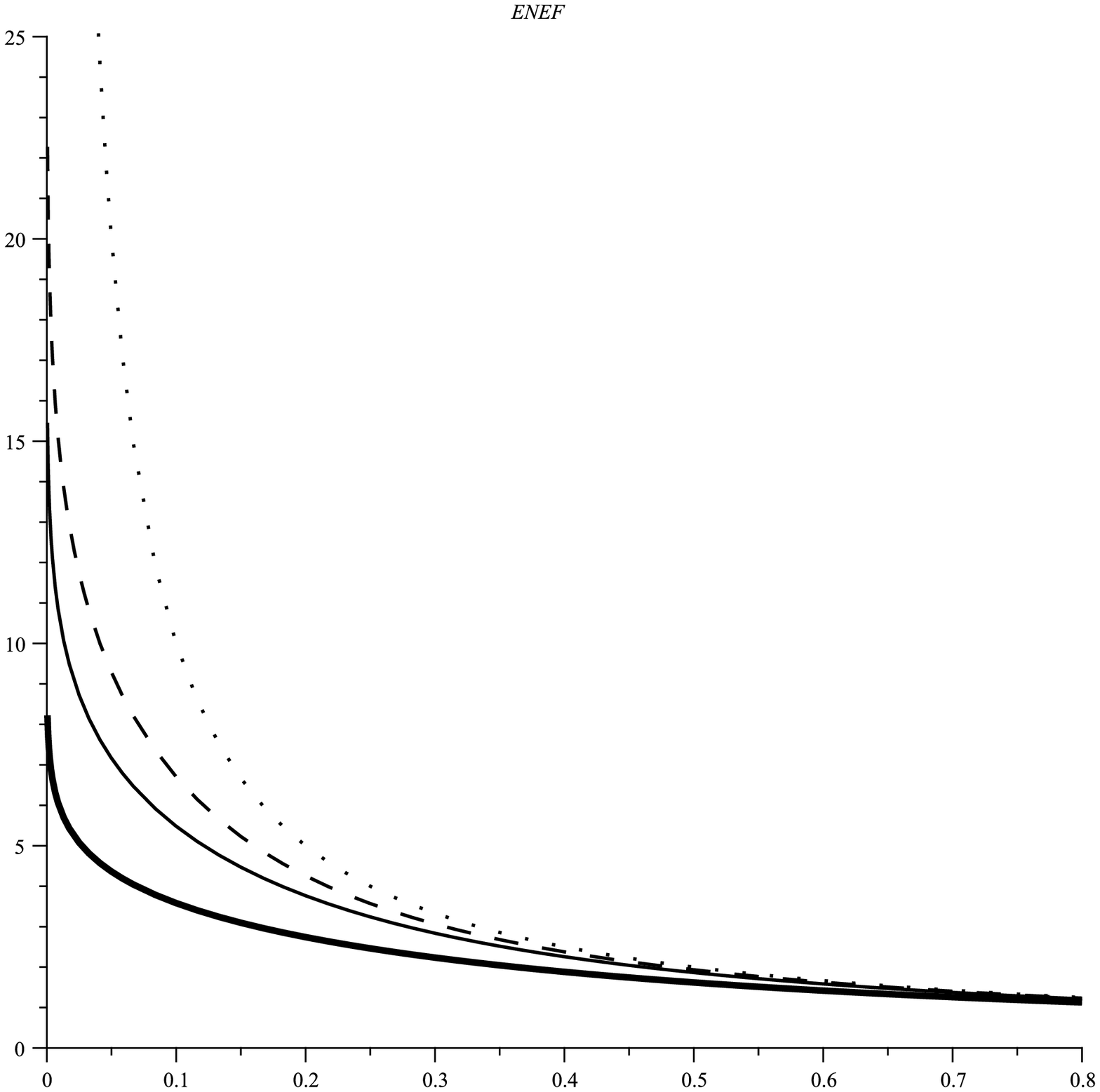} & \epsfxsize=5cm \epsffile{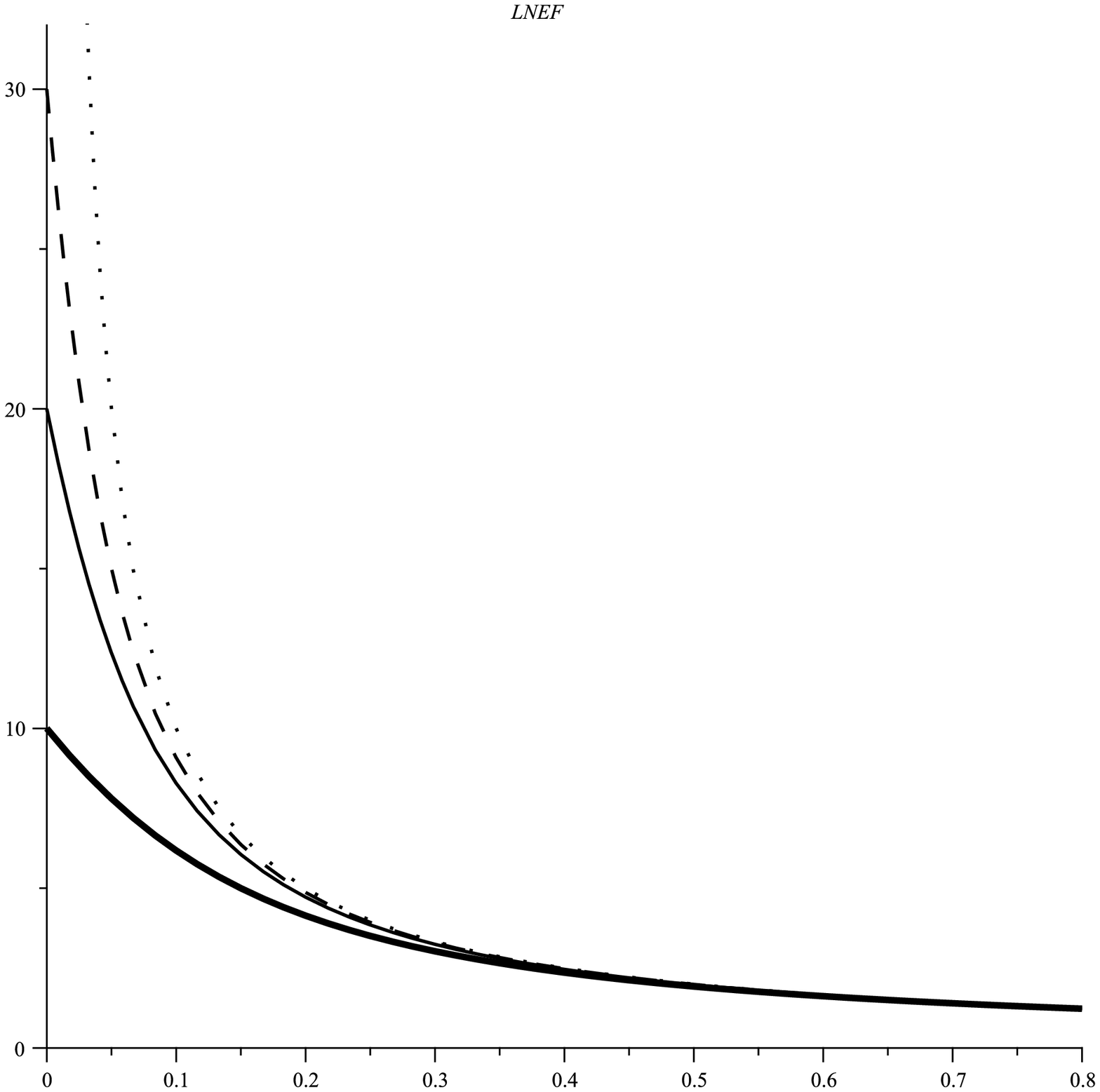}
\end{array}
$
\caption{$F_{tr}$ versus $r$ for $l=1$, $q=1$ and
$\protect\beta=5$ (Bold line), $\protect\beta=10$ (solid line),
$\protect\beta=15$ (dashed line) and $\protect\beta
\longrightarrow \infty$ (BTZ solution) (dotted line). "BINEF
(left), ENEF (middle) and LNEF (right)"} \label{ThirdE}
\end{figure}

\begin{figure}[tbp]
\epsfxsize=9cm \centerline{\epsffile{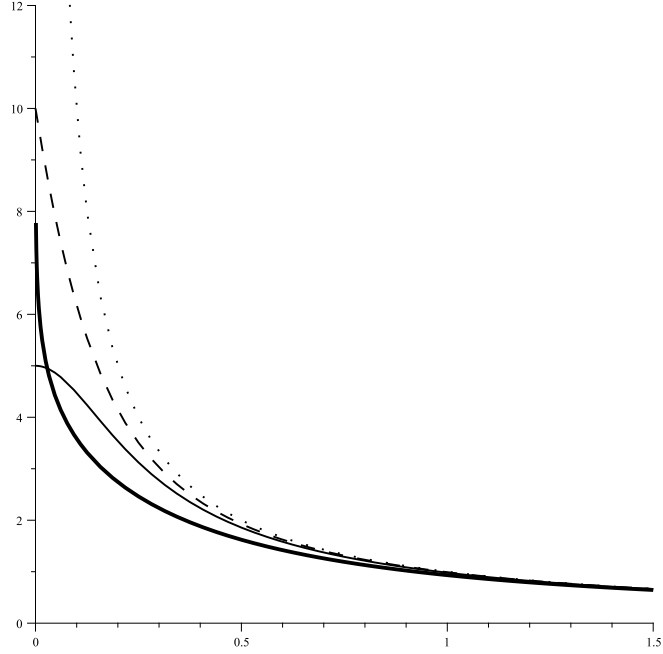}}
\caption{$F_{tr}$ versus $r$ in ENEF (Bold line), BINEF (solid
line), LNEF (dashed line) and BTZ solution (dotted line) for
$l=1$, $q=1$ and $\protect\beta=5$.} \label{Ftrthree}
\end{figure}

Considering Fig. \ref{ThirdE}, it is interesting to note that all three
types of the mentioned electromagnetic fields have finite values \emph{near}
the origin and they vanish at large values of $r$, as it should be. In
addition, this figure shows that the effect of nonlinearity parameter, $%
\beta $, on the strength of electromagnetic fields is more considerable for
small values of distances. Furthermore, Fig. \ref{Ftrthree} shows that the
mentioned NLED fields have different values for the fixed parameters and one
may think they have finite values at the origin ($r \rightarrow 0$), but it
is interesting to note that $F^{ENEF}_{tr}$ diverges at $r=0$ (see table A
for more details).

\begin{center}
\begin{tabular}{cccccccc}
\hline\hline
$r=$ & $10^{-2}$ & $10^{-10}$ & $10^{-20}$ & $10^{-50}$ & $10^{-100}$ & $%
10^{-1000}$ & $r\longrightarrow 0$ \\ \hline
$F_{tr}^{BINEF}$ & $4.993$ & $5.000$ & $5.000$ & $5.000$ & $5.000$ & $5.000$
& $5.000$ \\
$F_{tr}^{ENEF}$ & $5.941$ & $15.91$ & $23.16$ & $37.33$ & $53.18$ & $169.5$
& $\infty $ \\
$F_{tr}^{LNEF}$ & $9.512$ & $9.999$ & $10.00$ & $10.00$ & $10.00$ & $10.00$
& $10.00$ \\
$F_{tr}^{BTZ}$ & $10^{2}$ & $10^{10}$ & $10^{20}$ & $10^{50}$ & $10^{100}$ &
$10^{1000}$ & $\infty $ \\ \hline\hline
\end{tabular}
\\[0pt]
Table A: $F_{tr}$ for $\beta =5$, $q=1$, $l=1$ and small values of $r$.
\end{center}

%%%%%%%%%%%%%%%%%%%%%%%%%%%%%%%%%%%%%%%%%%%

To find the metric function $f(r)$, one may use any components of
Eq. (\ref{FE1}). Considering the function $h(r)$, the nontrivial
independent components of the field equation, (\ref{FE1}), are
\begin{equation}
\begin{array}{rr}
\left.
\begin{array}{r}
f^{\prime }(r)-\frac{2r}{l^{2}}-4r\beta ^{2}\left( 1-\Gamma \right) =0, \\
f^{\prime \prime }(r)-\frac{2}{l^{2}}-4\beta ^{2}\left( 1-\Gamma
^{-1}\right) =0.
\end{array}
\right\} , & \text{BINEF}\vspace{0.2cm} \\
\left.
\begin{array}{r}
f^{\prime }(r)-\frac{2r}{l^{2}}+r\beta ^{2}\left[ 1-\frac{\Gamma }{\sqrt{%
L_{W}}}\left( 1-L_{W}\right) \right] =0, \\
f^{\prime \prime }(r)-\frac{2}{l^{2}}+\beta ^{2}\left( 1-\frac{\Gamma }{%
\sqrt{L_{W}}}\right) =0.
\end{array}
\right\} , & \text{ENEF}\vspace{0.2cm} \\
\left.
\begin{array}{r}
f^{\prime }(r)-\frac{2r}{l^{2}}+8r\beta ^{2}\ln \left[ 1-\left(
\frac{\beta r(1-\Gamma )}{q}\right) ^{2}\right] +\frac{16\beta
^{4}r^{3}\left( 1-\Gamma \right) ^{2}}{q^{2}\left[ 1-\left(
\frac{\beta r(1-\Gamma )}{q}\right) ^{2}
\right] }=0, \\
f^{\prime \prime }(r)-\frac{2}{l^{2}}+8\beta ^{2}\ln \left( 1-\left( \frac{%
\beta r(1-\Gamma )}{q}\right) ^{2}\right) =0.%
\end{array}
\right\} , & \text{LNEF}%
\end{array}
.  \label{f(r)eq}
\end{equation}
The solutions of Eq. (\ref{f(r)eq}) can be written as
\begin{equation}
f(r)=\frac{r^{2}}{l^{2}}-M+\left\{
\begin{array}{ll}
2r^{2}\beta ^{2}\left( 1-\Gamma \right) +q^{2}\left[ 1-2\ln \left( \frac{%
r\left( 1+\Gamma \right) }{2l}\right) \right] , & \text{BINEF}\vspace{0.2cm}
\\
\frac{\beta rq\left( 1-2L_{W}\right) }{\sqrt{L_{W}}}-\frac{\beta ^{2}r^{2}}{2%
}+q^{2}\left[ \ln \left( \frac{\beta ^{2}l^{2}}{2q^{2}}\right) -{Ei}\left( 1,%
\frac{L_{W}}{2}\right) -\gamma +3\right] , & \text{ENEF}\vspace{0.2cm} \\
4\beta ^{2}r^{2}\left[ \ln \left( \frac{\Gamma +1}{2}\right) +3\right] -q^{2}%
\left[ \ln \left( \frac{\beta ^{2}r^{4}(\Gamma -1)(\Gamma +1)^{3}}{%
4q^{2}l^{2}}\right) +\frac{6}{\Gamma -1}-2\right] , & \text{LNEF}%
\end{array}
\right. ,  \label{F(r)}
\end{equation}
where $M$ is the integration constant which is related to mass parameter.

\subsection{Properties of the solutions}

It is easy to show that for the metric (\ref{Metric}), the Ricci and the
Kretschmann scalars are
\begin{eqnarray}
R &=&-f^{\prime \prime }(r)-\frac{2f^{\prime }(r)}{r}  \label{R} \\
R_{\mu \nu \rho \sigma }R^{\mu \nu \rho \sigma } &=&f^{\prime \prime
2}(r)+2\left( \frac{f^{\prime }(r)}{r}\right) ^{2},  \label{RR}
\end{eqnarray}
where prime and double primes denote first and second derivative with
respect to $r$, respectively. Also one can show that other curvature
invariants (such as Ricci square) are functions of $f^{\prime \prime }$ and $%
f^{\prime }/r$ and so it is sufficient to study the Ricci and the
Kretschmann scalars for the investigation of the spacetime curvature.
Considering Eq. (\ref{F(r)}), one can expand the Ricci and the Kretschmann
scalars near the origin
\begin{equation}
R=\left\{
\begin{array}{ll}
\frac{\zeta _{1}}{r^{2}}+O(r^{0}), & \text{BTZ} \\
\frac{\zeta _{2}}{r}+O(r^{0}), & \text{BINEF} \\
\frac{\zeta _{3}\sqrt{L_{W}}}{r}+\frac{\zeta
_{4}}{r\sqrt{L_{W}}}+O(r^{0}), & \text{ENEF} \\
\frac{\zeta _{5}}{r}+\zeta _{6}\ln r+O(r^{0}), & \text{LNEF}
\end{array}
\right.  \label{Rorigin}
\end{equation}
\begin{equation}
R_{\mu \nu \rho \sigma }R^{\mu \nu \rho \sigma}=\left\{
\begin{array}{ll}
\frac{\xi _{1}}{r^{4}}+\frac{\xi _{2}}{r^{2}}+O(r^{0}), & \text{BTZ} \\
\frac{\xi _{3}}{r^{2}}+\frac{\xi _{4}}{r}+O(r^{0}), & \text{BINEF} \\
\frac{\xi _{5}L_{W}}{r^{2}}+\frac{\xi _{6}}{r^{2}}+\frac{\xi _{7}}{r^{2}L_{W}%
}+\frac{\xi _{8}\sqrt{L_{W}}}{r}+\frac{\xi _{9}}{r\sqrt{L_{W}}}+O(r^{0}), &
\text{ENEF} \\
\frac{\xi _{10}}{r^{2}}+\frac{\xi _{11}\ln r}{r}+\frac{\xi
_{12}}{r}+\xi _{13}(\ln r)^{2}+\xi _{14}\ln r+O(r^{0}), &
\text{LNEF}
\end{array}%
\right.  \label{RRorigin}
\end{equation}
where $\zeta _{i}$'s and $\xi _{i}$'s are functions of $\beta $, $l$ and $q$%
. We should note that the functions $L_{W}$ and $\sqrt{L_{W}}$ go to
infinity for $r\longrightarrow 0$, but much weaker than $r^{-2}$ and $r^{-1}$%
. So it is easy to find that
\begin{eqnarray}
\lim_{r\longrightarrow 0^{+}}R &=&\infty , \\
\lim_{r\longrightarrow 0^{+}}R_{\mu \nu \rho \sigma }R^{\mu \nu \rho \sigma
} &=&\infty ,
\end{eqnarray}
which confirm that, like BTZ black hole, the metric given by Eqs.
(\ref{Metric}) and (\ref{F(r)}) has an essential timelike
singularity at $r=0$. We should note that the singularity strength
is different for BTZ, BINEF, ENEF and LNEF solutions and also the
nonlinearity of electromagnetic field reduces the strength of
singularity.

In order to consider the asymptotic behavior of the solution, we calculate
the Ricci and the Kretschmann scalars for large values of $r$
\begin{eqnarray}
\lim_{r\longrightarrow \infty }R &=&-\frac{6}{l^{2}}+\frac{2q^{2}}{r^{2}}+%
\frac{\chi q^{4}}{4\beta ^{2}r^{4}}+O\left( \frac{1}{r^{6}}\right) ,
\label{Rinf} \\
\lim_{r\longrightarrow \infty }R_{\mu \nu \rho \sigma }R^{\mu \nu \rho
\sigma } &=&\frac{12}{l^{4}}-\frac{8q^{2}}{r^{2}l^{2}}+\frac{q^{4}\left(
12\beta ^{2}l^{2}-\chi \right) }{\beta ^{2}l^{2}r^{4}}+O\left( \frac{1}{r^{6}%
}\right) ,  \label{RRinf}
\end{eqnarray}%
where $\chi =2$, $8$ and $1$ for BINEF, ENEF and LNEF, respectively.
Equations (\ref{Rinf}) and (\ref{RRinf}) confirm that the asymptotic
behavior of the obtained solutions is AdS.
\begin{figure}[tbp]
$%
\begin{array}{cc}
\epsfxsize=7cm \epsffile{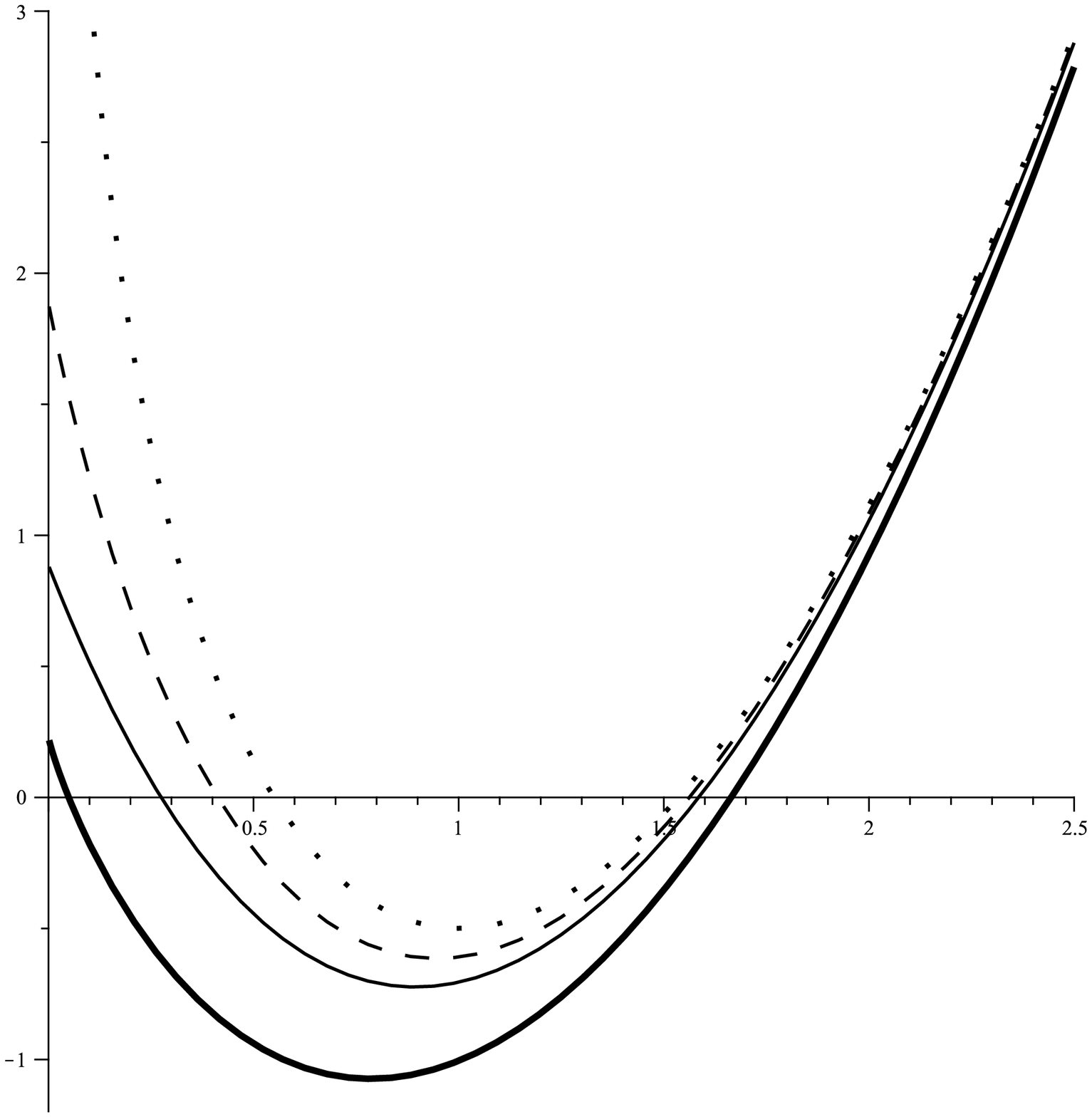} & \epsfxsize=7cm \epsffile{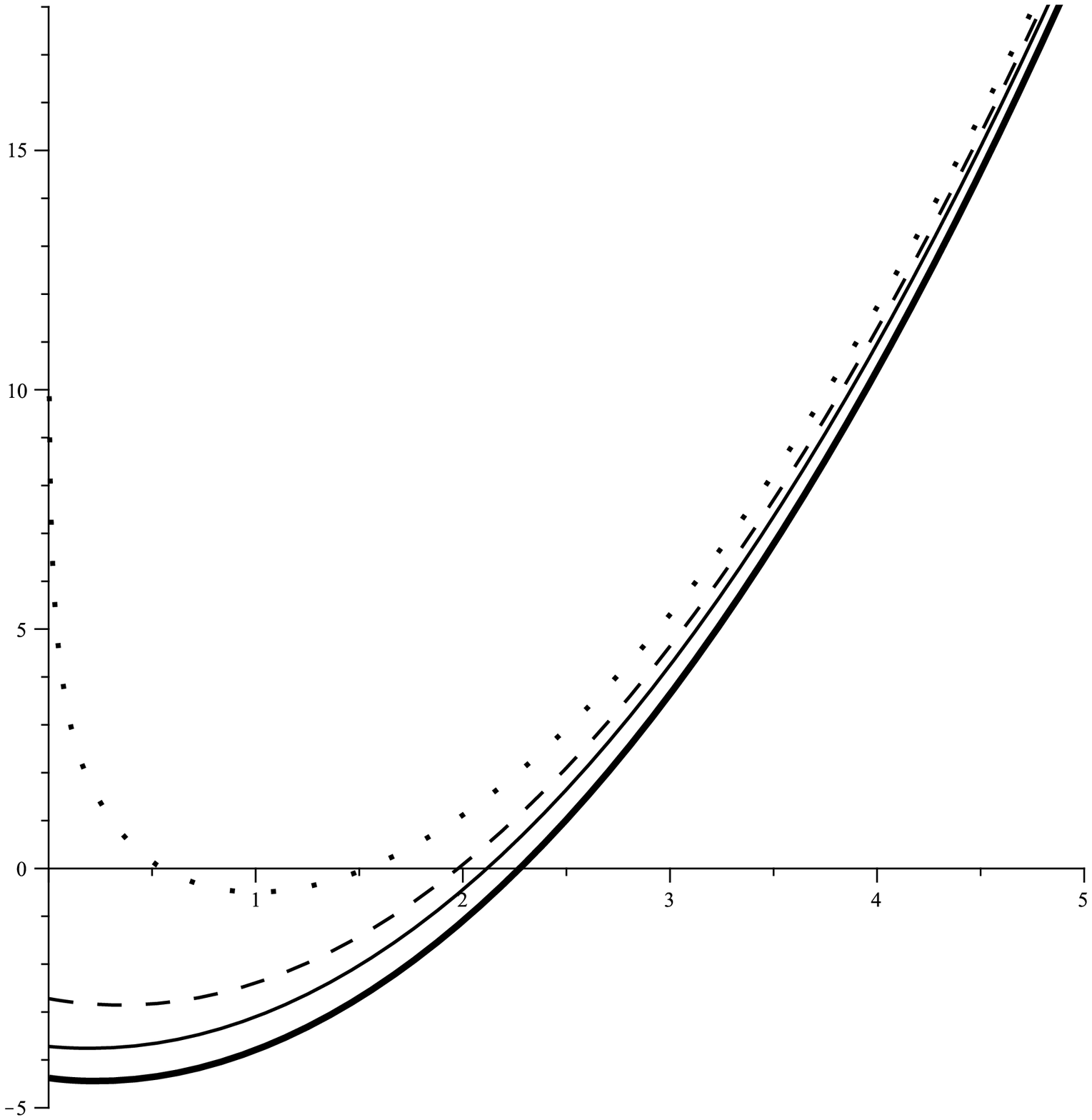}%
\end{array}
$%
\caption{$f(r)$ versus $r$ for $l=1$, $q=1$, $M=1.5$, and
$\protect\beta =1$ (two horizons: left figure) and $\protect\beta
=0.1$ (one horizon: right figure) in ENEF branch (Bold line),
BINEF branch (solid line), LNEF branch (dashed line) and BTZ
solution (dotted line).} \label{OneTwohorizon}
\end{figure}
\begin{figure}[tbp]
$
\begin{array}{ccc}
\epsfxsize=5cm \epsffile{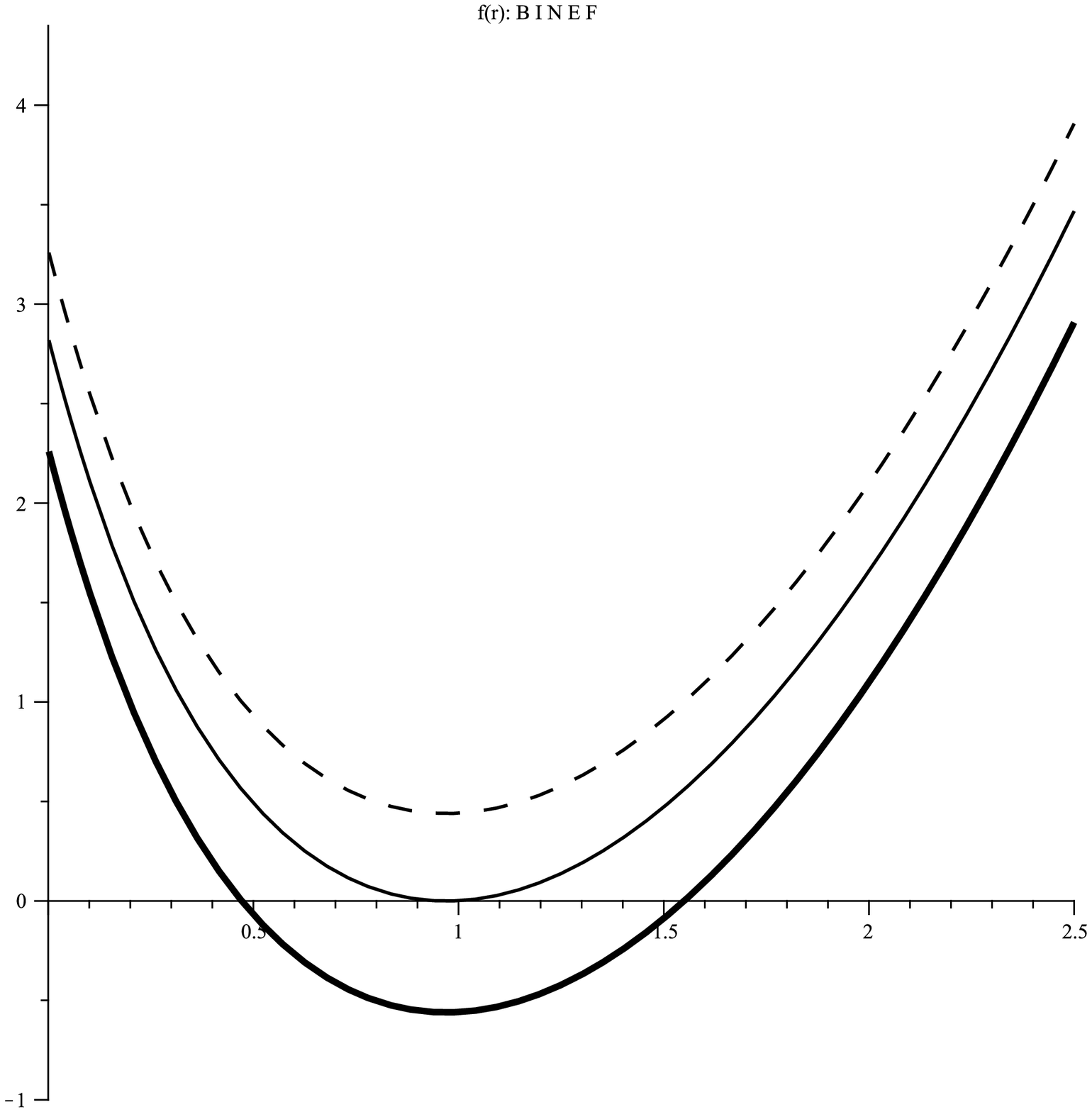} & \epsfxsize=5cm
\epsffile{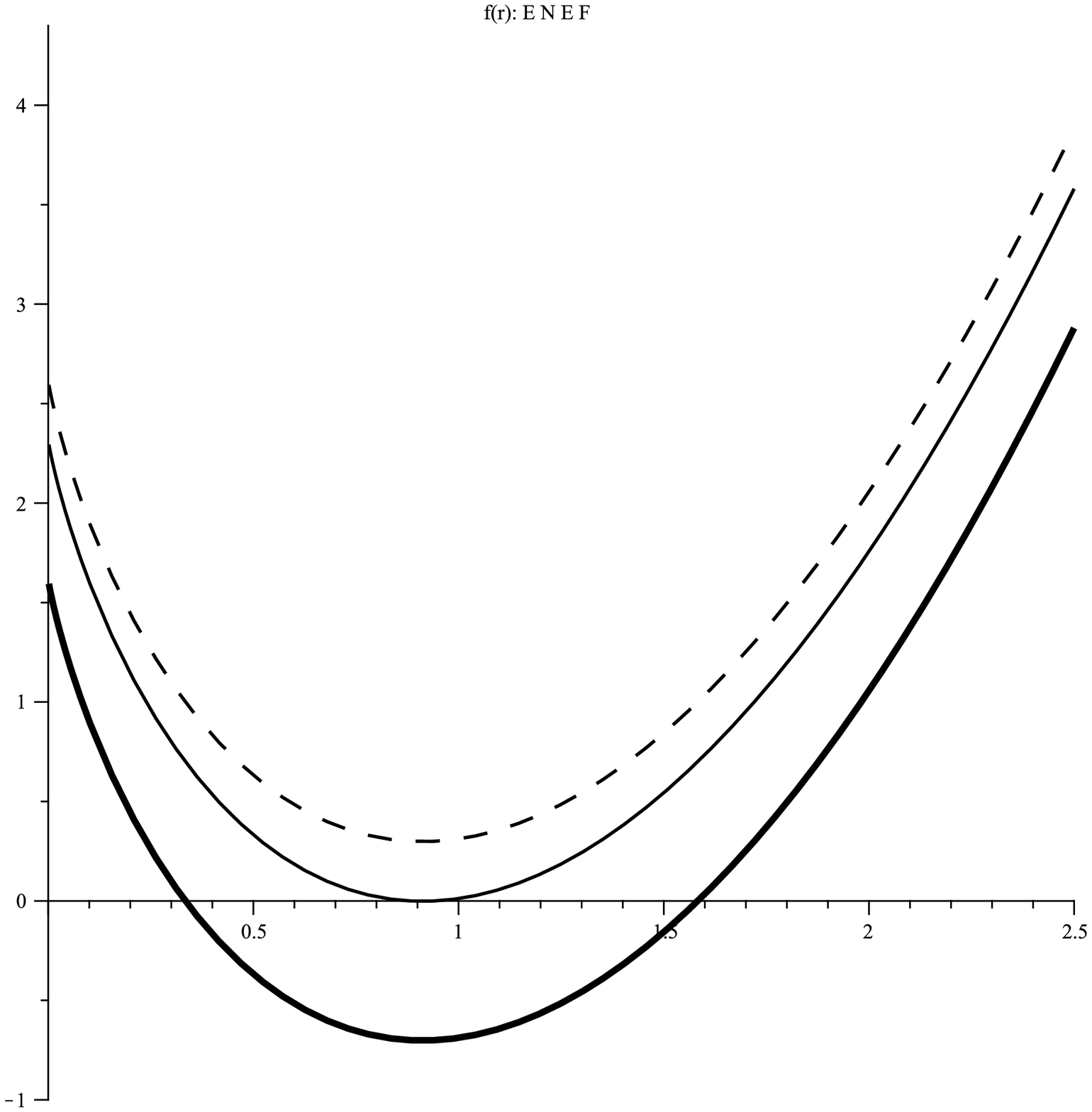} & \epsfxsize=5cm \epsffile{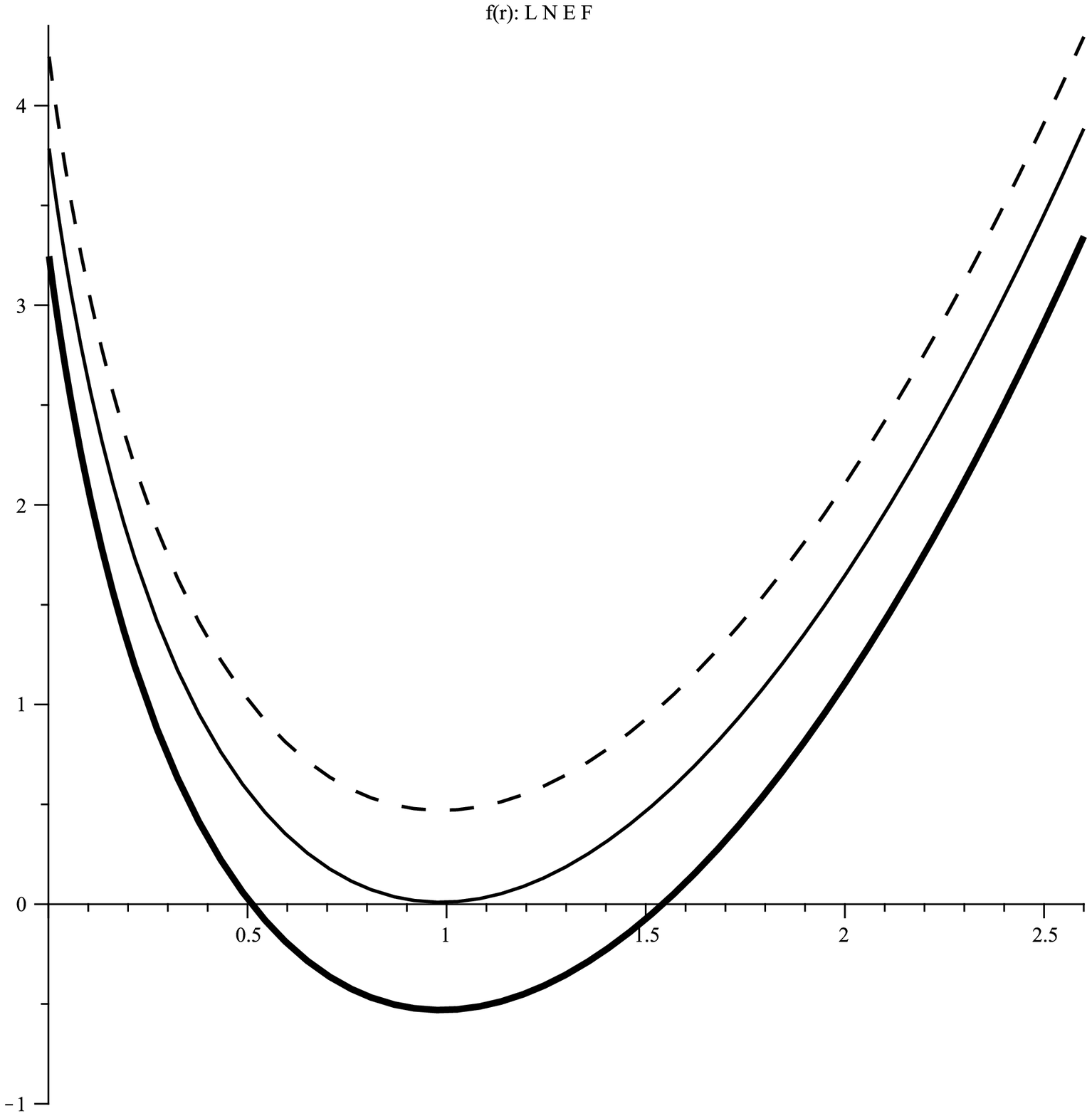}
\end{array}
$
\caption{$f(r)$ versus $r$ for $l=1$, $q=1$, $\protect\beta =2$
and $M>M_{ext}$ (Bold line), $M=M_{ext}$ (solid line)
and$M<M_{ext}$ (dashed line), where $M_{ext}=0.94$ for BINEF,
$M_{ext}=0.8$ for ENEF and $M_{ext}=0.96$ for LNEF. "BINEF (left),
ENEF (middle) and LNEF (right)"} \label{Thirdf}
\end{figure}
In addition, it is easy to find that the laps function of charged
BTZ black hole is positive for both $r\longrightarrow 0$ and
$r\longrightarrow \infty $ and therefore depend on the values of
the metric parameters, one can obtain a black hole with two
horizons, an extreme black hole and a naked singularity. But for
the nonlinear charged solutions, (Eq. (\ref{F(r)})), we should not
that depends on the value of the nonlinearity parameter, $\beta $,
the function $f(r)$ is zero, positive or negative near the origin
(see Figs. \ref{OneTwohorizon} and \ref{Thirdf} for more details).
In other word, one can find that for $r\longrightarrow \infty $,
the lapse function (Eq. (\ref{F(r)})) is positive but near the
origin ($r\longrightarrow 0$), the lapse function (Eq.
(\ref{F(r)})) may be positive, zero and negative for $\beta
>\beta _{c}$ , $\beta =\beta _{c}$ and $\beta <\beta _{c}$,
respectively. Considering $\lim_{r\longrightarrow 0^{+}}f(r)=0$,
we may find $\beta _{c}$ as a function of $l$, $q$ and $M$. It is
so interesting to note that, unlike charged BTZ black hole, the
metric function of nonlinear charged black hole solutions (Eq.
(\ref{F(r)})) behaves like as uncharged (Schwarzschild) solution
(Fig. \ref{OneTwohorizon}, right), for $\beta <\beta _{c}$. In
other word, for $\beta <\beta _{c}$, the function $f(r)$ has one
real positive root at $r=r_{+}$, where $f^{\prime }(r=r_{+})\neq
0$.

\subsection{Conserved and thermodynamics quantities \label{Therm}}

The Hawking temperature of the black hole on the outer horizon $r_{+}$, may
be obtained through the use of the definition of surface gravity,
\[
T_{+}=\beta _{+}^{-1}=\frac{1}{2\pi }\sqrt{-\frac{1}{2}\left( \nabla _{\mu
}\chi _{\nu }\right) \left( \nabla ^{\mu }\chi ^{\nu }\right) }
\]%
where $\chi =\partial /\partial t$ is the Killing vector. One obtains
\begin{equation}
T_{+}=\left\{
\begin{array}{ll}
\frac{r_{+}}{2l^{2}\Gamma _{+}}+\frac{q^{2}\left( \Gamma _{+}^{-1}-2\beta
^{2}l^{2}\right) }{2r_{+}\beta ^{2}l^{2}\left( 1+\Gamma _{+}\right) }, &
\text{BINEF}\vspace{0.2cm} \\
\frac{1}{4}\left[ r_{+}\left( \frac{2}{l^{2}}-\beta ^{2}\right) +2q\beta
\sqrt{L_{W_{+}}}\left( \frac{1}{L_{W_{+}}}-1\right) \right] , & \text{ENEF}%
\vspace{0.2cm} \\
\frac{r_{+}}{2l^{2}}+\frac{2q^{2}\ln \left( \frac{\exp (1)\left( 1+\Gamma
_{+}\right) }{2}\right) }{r_{+}\Gamma _{+}\left( \Gamma _{+}-1\right) }-%
\frac{4q^{2}+4\beta ^{2}r_{+}^{2}\ln \left( \frac{\exp (2)\left( 1+\Gamma
_{+}\right) }{2}\right) }{2r_{+}\Gamma _{+}}, & \text{LNEF}%
\end{array}%
\right.  \label{Temp}
\end{equation}%
where $\Gamma _{+}=\sqrt{1+\frac{q^{2}}{r_{+}^{2}\beta ^{2}}}$ and $%
L_{W_{+}}=LambertW(\frac{4q^{2}}{\beta ^{2}r_{+}^{2}})$.

The electric potential $U$, measured at infinity with respect to the
horizon, is defined by \cite{Gub}
\begin{eqnarray}
U &=&A_{\mu }\chi ^{\mu }\left\vert _{reference}-A_{\mu }\chi ^{\mu
}\right\vert _{r=r_{+}}= \\
&&-\frac{q}{2}\times \left\{
\begin{array}{ll}
2\ln \left[ \frac{r_{+}}{2l}\left( 1+\Gamma _{+}\right) \right] , & \text{%
BINEF}\vspace{0.1cm} \\
\frac{\sqrt{L_{W_{+}}}\beta r_{+}}{q}+{Ei}\left( 1,\frac{L_{W_{+}}}{2}%
\right) +\ln \left( \frac{2q^{2}}{l^{2}\beta ^{2}}\right) +\gamma -2, &
\text{ENEF}\vspace{0.1cm} \\
2\ln \left[ \frac{r_{+}}{2l}\left( 1+\Gamma _{+}\right) \right] -\frac{%
2r_{+}^{2}\beta ^{2}}{q^{2}}\left( 1-\Gamma _{+}\right) -1, & \text{LNEF}%
\end{array}%
\right. .  \label{U}
\end{eqnarray}%
where in the reference, $U$ should vanish.

More than thirty years ago, Bekenstein argued that the entropy of
a black hole is a linear function of the area of its event
horizon, which so called area law \cite{Bekenstein}. Since the
area law of the entropy is universal, and applies to all kinds of
black holes in Einstein gravity \cite{Bekenstein,Hawking3},
therefore the entropy of the obtained charged black hole solutions
is equal to one-quarter of the area of the horizon, i.e.,
\begin{equation}
S=\frac{\pi r_{+}}{2}.  \label{Entropy}
\end{equation}
The electric charge of the black holes, $Q$, can be found by calculating the
flux of the electromagnetic field at infinity, yielding
\begin{equation}
Q=\frac{\pi q}{2},  \label{Charg}
\end{equation}
for all three types of the mentioned NLED fields.

The present spacetime (\ref{Metric}), have boundary with timelike ($\xi
=\partial /\partial t$) Killing vector field. It is straightforward to show
that for the quasi-local mass, we can write
\begin{equation}
\mathcal{M}=\int_{\mathcal{B}}d\varphi \sqrt{\sigma }T_{ab}n^{a}\xi ^{b}=%
\frac{\pi }{8}M.  \label{Mas}
\end{equation}

Here, we check the first law of thermodynamics for our solutions. We obtain
the mass as a function of the extensive quantities $S$ and $Q$. One may then
regard the parameters $S$, and $Q$ as a complete set of extensive parameters
for the mass $\mathcal{M}(S,Q)$
\begin{equation}
\mathcal{M}(S,Q)=\frac{S^{2}}{2\pi l^{2}}+\left\{
\begin{array}{ll}
\frac{\beta ^{2}S^{2}\left( 1-\Upsilon \right) }{\pi }+\frac{4Q^{2}\left[
1-2\ln \left( \frac{S\left( \Upsilon +1\right) }{\pi l}\right) \right] }{%
2\pi }, & \text{BINEF}\vspace{0.2cm} \\
\frac{\beta SQ}{2\pi \sqrt{\Pi }}-\frac{\beta ^{2}S^{2}}{4\pi }+\frac{Q^{2}%
\left[ 3-\gamma -{EI}-\frac{2\beta S\sqrt{\Pi }}{Q}+\ln \left( \frac{\pi
^{2}\beta ^{2}l^{2}}{8Q^{2}}\right) \right] }{2\pi }, & \text{ENEF}\vspace{%
0.2cm} \\
\frac{2\beta ^{2}S^{2}\left[ 3+\ln \left( \frac{\Upsilon +1}{2}\right) %
\right] }{\pi }+\frac{Q^{2}\left( 1-\frac{3}{\Upsilon -1}-\ln \left[ \frac{%
\beta S^{2}\sqrt{\Upsilon ^{2}-1}}{\pi lQ(\Upsilon +1)^{-1}}\right] \right)
}{\pi }, & \text{LNEF}%
\end{array}
\right. ,  \label{Msmar}
\end{equation}
where $\Upsilon =\sqrt{1+\left( \frac{Q}{S\beta }\right) ^{2}}$ ,
$EI={Ei} \left( 1,\frac{\Pi }{2}\right) $ and $\Pi
=LambertW(\frac{4Q^{2}}{\beta ^{2}S^{2}})$. Now, we should
differentiate $\mathcal{M}(S,Q)$ to obtain
\begin{equation}
d\mathcal{M}(S,Q)=\left( \frac{\partial \mathcal{M}}{\partial S}\right)
_{Q}dS+\left( \frac{\partial \mathcal{M}}{\partial Q}\right) _{S}dQ,
\label{dM}
\end{equation}
where
\begin{equation}
\left( \frac{\partial \mathcal{M}}{\partial S}\right) _{Q}=\left\{
\begin{array}{ll}
\frac{\left( \Upsilon +1\right) +\frac{Q^{2}}{\beta ^{2}S^{2}}\left(
1-2l^{2}\beta ^{2}\Upsilon \right) }{\frac{\pi l^{2}}{S}\Upsilon \left(
1+\Upsilon \right) }, & \text{BINEF}\vspace{0.2cm} \\
-\frac{2\beta Ql^{2}\left( \Pi -1\right) +S\sqrt{\Pi }\left[ l^{2}\beta
^{2}-2\right] }{2\pi l^{2}\sqrt{\Pi }}, & \text{ENEF}\vspace{0.2cm} \\
\frac{4S\beta ^{2}\left[ \Upsilon ^{2}-\Upsilon \right] \left[ \ln
\left( \frac{\Upsilon }{2}\right) +\frac{1}{4l^{2}\beta
^{2}}+2\right] }{\pi \Upsilon \left( \Upsilon -1\right) }, &
\text{LNEF}
\end{array}
\right. ,  \label{dMS}
\end{equation}
and
\begin{equation}
\left( \frac{\partial \mathcal{M}}{\partial Q}\right) _{S}=\left\{
\begin{array}{ll}
\frac{-2Q\left( \Upsilon ^{2}+\Upsilon \right) \ln \left[
\frac{S}{\pi l} \left( 1+\Upsilon \right) \right] }{\pi \Upsilon
\left( 1+\Upsilon \right) },
& \text{BINEF}\vspace{0.2cm} \\
\frac{Q\left[ 2-\gamma -EI+\ln \left( \frac{\pi ^{2}\beta
^{2}l^{2}}{8Q^{2}}
\right) \right] -\beta S\sqrt{\Pi }}{\pi }, & \text{ENEF}\vspace{0.2cm} \\
\frac{2Q^{3}\left[ 1+\frac{2S^{2}\beta ^{2}}{Q^{2}}-\ln \left(
\frac{\beta S^{2}\left( \Upsilon +1\right) \sqrt{\Upsilon
^{2}-1}}{\pi lQ}\right) -\frac{1}{\left( \Upsilon -1\right)
}\right] }{\pi S^{2}\beta ^{2}\left( \Upsilon ^{2}-1\right) }, &
\text{LNEF}
\end{array}
\right. ,  \label{dMQ}
\end{equation}
At this point, we should replace $S$ and $Q$ from Eqs.
(\ref{Entropy}) and (\ref{Charg}), and rewrite $\left(
\frac{\partial \mathcal{M}}{\partial S}\right) _{Q}$ and $\left(
\frac{\partial \mathcal{M}}{\partial Q}\right) _{S}$ which are the
same as Eqs. (\ref{Temp}) and (\ref{U}), respectively. In other
word, we could define the intensive parameters conjugate to
extensive quantities $S$ and $Q$. These quantities are the
temperature and the electric potential
\begin{equation}
T=\left( \frac{\partial \mathcal{M}}{\partial S}\right) _{Q},\ \ U=\left(
\frac{\partial \mathcal{M}}{\partial Q}\right) _{S},  \label{Dsmar}
\end{equation}%
where the intensive quantities calculated by Eq. (\ref{Dsmar}) coincide with
Eqs. (\ref{Temp}) and (\ref{U}). Thus, these quantities satisfy the first
law of thermodynamics
\begin{equation}
d\mathcal{M}=TdS+UdQ.  \label{1thLaw}
\end{equation}

\subsection{Thermodynamic stability in the canonical ensemble}

Now, we investigate the thermodynamic stability of nonlinear
charged black hole solutions in the canonical ensemble. The
stability of a thermodynamic system with respect to the small
variations of the thermodynamic coordinates, can be studied in the
canonical ensemble which the charge is fixed parameter. In other
word, the positivity of the heat capacity $C_{Q}=T_{+}/(\partial
^{2}\mathcal{M}/\partial S^{2})_{Q}$ is sufficient to ensure the
local stability. It is straightforward to show that $(\partial
^{2}\mathcal{M}/\partial S^{2})_{Q}$ is
\begin{equation}
\left( \frac{\partial ^{2}\mathcal{M}}{\partial S^{2}}\right)
_{Q}=\frac{1}{\pi l^{2}}+\left\{
\begin{array}{ll}
\frac{2q^{2}}{\pi r^{2}\left( 1+\Gamma _{+}\right) \Gamma _{+}}, & \text{
BINEF}\vspace{0.1cm} \\
\frac{\beta ^{2}}{2\pi }\left( e^{\frac{L_{W_{+}}}{2}}-1\right) , & \text{
ENEF}\vspace{0.1cm} \\
\frac{4\beta ^{2}}{\pi }\ln \left( \frac{1+\Gamma _{+}}{2}\right)
, & \text{ LNEF}
\end{array}
\right. .  \label{dMSS1}
\end{equation}
\begin{figure}[tbp]
\epsfxsize=9cm \centerline{\epsffile{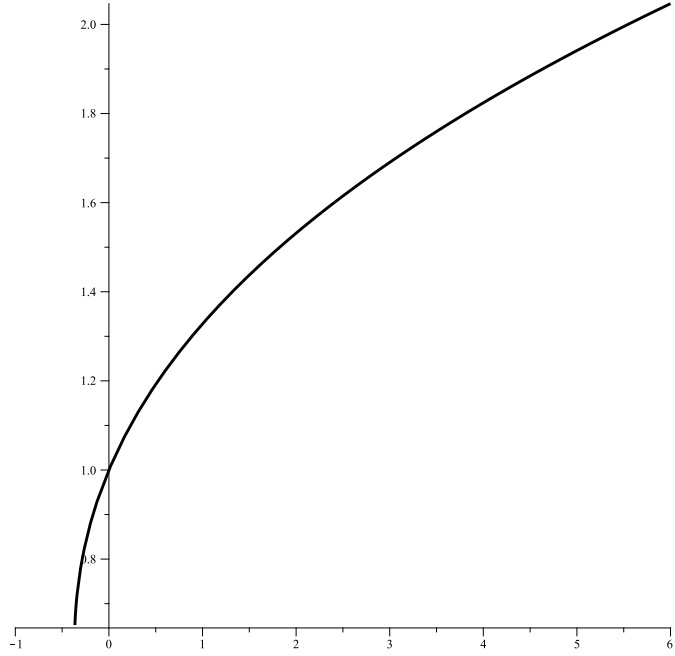}}
\caption{$e^{\frac{LambertW(x)}{2}}$ versus $x$, which shows that
$e^{\frac{LambertW(x)}{2}} \geq 1$ for $x \geq 0$.}
\label{dMSSfig}
\end{figure}
It is clear to find that $(\partial ^{2}\mathcal{M}/\partial
S^{2})_{Q}$ is positive for BINEF and LNEF branches. Considering
ENEF branch, one may find that $e^{\frac{L_{W_{+}}}{2}}\geq 1$ and
therefore $(\partial ^{2}\mathcal{M}/\partial S^{2})_{Q}$ is
positive (see Fig. \ref{dMSSfig} for more details). Since the
temperature of the black hole solutions is positive, it is clear
that the black holes are stable in the canonical ensemble.

\section{Asymptotic BTZ solutions}

Here we would like to find that for large distance ($r>>1$), the obtained
solutions are the same as charged BTZ solution. To do this, firstly, we
expand Eq. (\ref{h(r)}) for the large values of $r$ which leads to
\begin{equation}
\left. h(r)\right\vert _{\text{Large }r}=q\ln \left( \frac{r}{l}\right)
+\left\{
\begin{array}{ll}
\frac{q^{3}}{4r^{2}\beta ^{2}}-\frac{3q^{5}}{32r^{4}\beta ^{4}}+O\left(
\frac{1}{r^{6}}\right) , & \text{BINEF}\vspace{0.1cm} \\
\frac{q^{3}}{r^{2}\beta ^{2}}-\frac{5q^{5}}{2r^{4}\beta
^{4}}+O\left( \frac{1}{r^{6}}\right) , & \text{ENEF}\vspace{0.1cm} \\
\frac{q^{3}}{8r^{2}\beta ^{2}}-\frac{q^{5}}{32r^{4}\beta
^{4}}+O\left( \frac{1}{r^{6}}\right) , & \text{LNEF}
\end{array}
\right. .  \label{asymph}
\end{equation}
It is easy to find that for large $r$, the dominant (first) term
of $h(r)$ for all the mentioned NLED fields is the same as one in
BTZ solution \cite{BTZ1,BTZ2}. Differentiating from Eq.
(\ref{asymph}) or expanding Eq. (\ref{Ftr}) for large distance,
one can obtain
\begin{equation}
F_{tr}=\frac{q}{r}+\left\{
\begin{array}{ll}
-\frac{q^{3}}{2\beta ^{2}r^{3}}+\frac{3q^{5}}{8\beta ^{4}r^{5}}+O\left(
\frac{1}{r^{7}}\right) , & \text{BINEF} \\
-\frac{2q^{3}}{\beta ^{2}r^{3}}+\frac{10q^{5}}{\beta ^{4}r^{5}}+O\left(
\frac{1}{r^{7}}\right) , & \text{ENEF} \\
-\frac{q^{3}}{4\beta ^{2}r^{3}}+\frac{q^{5}}{8\beta
^{4}r^{5}}+O\left( \frac{1}{r^{7}}\right) , & \text{LNEF}
\end{array}
\right. ,  \label{asympFtr}
\end{equation}
which its dominant (first) term is similar to the electrical field
of $(2+1)$-dimensional Reissner-Nordstr\"{o}m black hole.

Now, we focus on the metric function. Straightforward calculations show that
expansion of $f(r)$ in Eq. (\ref{F(r)}) for $r>>1$, leads to the following
equation
\begin{equation}
f(r)=\frac{r^{2}}{l^{2}}-M-2q^{2}\ln \left( \frac{r}{l}\right) +\left\{
\begin{array}{ll}
-\frac{q^{4}}{4r^{2}\beta ^{2}}+\frac{q^{6}}{16r^{4}\beta ^{4}}+O\left(
\frac{1}{r^{6}}\right) , & \text{BINEF}\vspace{0.2cm} \\
-\frac{q^{4}}{r^{2}\beta ^{2}}+\frac{5q^{6}}{3r^{4}\beta
^{4}}+O\left( \frac{1}{r^{6}}\right) , & \text{ENEF}\vspace{0.2cm} \\
-\frac{q^{4}}{8r^{2}\beta ^{2}}+\frac{q^{6}}{48r^{4}\beta
^{4}}+O\left( \frac{1}{r^{6}}\right) , & \text{LNEF}
\end{array}
\right. .  \label{asympF}
\end{equation}
One may ignore small charge terms to find laps function of
$(2+1)$-dimensional BTZ black hole.

\section{Conclusions}

Considering the nonlinear electromagnetic fields in various sciences is one
of the interests but with cumbersome calculations. In this paper, we
introduced three kinds of NLED theories which in the weak field
approximation (large values of nonlinearity parameter: $\beta
\longrightarrow \infty $) become the usual linear Maxwell theory. In
addition, as we have shown, presented solutions have the following
properties:

First, obtained electromagnetic fields have regular behavior for large
values of distance and they are finite near the origin. In addition, we
showed that the effect of nonlinearity parameter, on the strength of
electromagnetic fields is more considerable for small values of distances.
It is so interesting that $F_{tr}^{ENEF}$ diverges at $r=0$, but its
divergency is very slower than the electromagnetic field of BTZ solution.

Second, all obtained solutions have a timelike curvature
singularity at $r=0$, and they are asymptotic AdS. In other word,
the NLED fields have no effect on the existence of singularity and
asymptotic behavior, but we should note that the nonlinearity of
electromagnetic field reduces the strength of singularity.
Furthermore, for small values of the nonlinearity parameter,
($\beta < \beta_{c}$), the singularity covered with a non-extreme
horizon. In other word, in this case the horizon geometry of
nonlinear charged black holes is close to the horizon of uncharged
(Schwarzschild) black hole solution.

Third, obtained solutions have different temperature and electric potential,
but the same entropy and electric charge. We should note that, unlike the
solutions of Einstein-power Maxwell invariant gravity \cite{PMIpaper}, the
electric charge is the same as Maxwell field and the nonlinearity has no
effect on it.

Fourth, one may confirm that all obtained solutions reduce to the charged
BTZ black hole for large values of distance.

As final remarks, we should note that the conserved and thermodynamic
quantities satisfy the first law of thermodynamics and the presented
solutions have positive heat capacity, which means that the black holes are
stable for all the allowed values of the metric parameters.

It is worthwhile to investigate the dynamical stability of the presented
solutions and also generalize these three dimensional solutions to higher
dimensional case with various horizon topologies, and these problems are
left for the future.


\begin{thebibliography}{999}
\bibitem{Fradkin85} E. S. Fradkin and A. A. Tseytlin, Phys. Lett. B 163, 123
(1985);

E. Bergshoeff, E. Sezgin, C. N. Pope and P. K. Townsend, Phys. Lett. B 188,
70 (1987);

R. R. Metsaev, M. A. Rahmanov and A. A. Tseytlin, Phys. Lett. B 193, 207
(1987);

A. A. Tseytlin, Nucl. Phys. B 501, 41 (1997);

D. Brecher and M. J. Perry, Nucl. Phys. B 527, 121 (1998).

\bibitem{BI} M. Born and L. Infeld, Proc. R. Soc. London A 143, 410 (1934);

M. Born and L. Infeld, Proc. R. Soc. London A 144, 425 (1934).

\bibitem{Leigh89} R. G. Leigh, Mod. Phys. Lett. A 4, 2767 (1989).

\bibitem{Hoffmann} B. Hoffmann, Phys. Rev. 47, 877 (1935).

\bibitem{BIpaper} M. H. Dehghani and H. R. Sedehi, Phys. Rev. D 74, 124018
(2006);

D. L. Wiltshire, Phys. Rev. D 38, 2445 (1988);

M. Aiello, R. Ferraro and G. Giribet, Phys. Rev. D 70, 104014 (2004);

M. H. Dehghani and S. H. Hendi, Int. J. Mod. Phys. D 16, 1829 (2007);

M. H. Dehghani, S. H. Hendi, A. Sheykhi and H. R. Rastegar-Sedehi, JCAP, 02,
020 (2007);

M. H. Dehghani, N. Alinejadi and S. H. Hendi, Phys. Rev. D 77, 104025 (2008);

S. H. Hendi, J. Math. Phys. 49, 082501 (2008).

\bibitem{PMIpaper} M. Hassaine and C. Martinez, Phys. Rev. D 75, 027502
(2007);

S. H. Hendi and H. R. Rastegar-Sedehi, Gen. Relativ. Gravit. 41,
1355 (2009);

S. H. Hendi, Phys. Lett. B 677, 123 (2009);

M. Hassaine and C. Martinez, Class. Quantum Gravit. 25, 195023 (2008);

H. Maeda, M. Hassaine and C. Martinez, Phys. Rev. D 79, 044012 (2009);

S. H. Hendi and B. Eslam Panah, Phys. Lett. B 684, 77 (2010);

S. H. Hendi, Phys. Lett. B 690, 220 (2010);

S. H. Hendi, Prog. Theor. Phys. 124, 493 (2010);

S. H. Hendi, Eur. Phys. J. C 69, 281 (2010);

S. H. Hendi, Phys. Rev. D 82, 064040 (2010).

\bibitem{Ayon98} E. Ayon-Beato and A. Garcia, Phys. Rev. Lett. 80, 5056
(1998).

\bibitem{Oliveira94} H. P. de Oliveira, Class. Quantum Gravit. 11, 1469 (1994).

\bibitem{Soleng} H. H. Soleng, Phys. Rev. D 52, 6178 (1995).

\bibitem{Oliv94} H. P. Oliveira, Class. Quantum Gravit. 11, 1469 (1994).

\bibitem{Palatnik98} D. Palatnik, Phys. Lett. B 432, 287 (1998).

\bibitem{Ayon46} E. Ayon--Beato and A. Garcia, Phys. Rev. Lett. 80, 5056
(1998).

\bibitem{Ayon84} E. Ayon--Beato and A. Garcia, Gen. Relativ. Gravit. 31, 629
(1999).

\bibitem{Ayon99} E. Ayon--Beato and A. Garcia, Phys. Lett. B 464, 25 (1999).

\bibitem{Rasheed97} D. A. Rasheed, [hep-th/9702087].

\bibitem{Boillat} G. Boillat, J. Math. Phys. 11, 941 (1970); 11, 1482 (1970).

\bibitem{GibRash} G. W. Gibbons and D. A. Rasheed, Nucl. Phys. B 454, 185
(1995).

\bibitem{Gopakumar} R. Gopakumar, S. Minwalla, N. Seiberg, A. Strominger,
JHEP 0008, 008 (2000).

\bibitem{Tamaki2000} T. Tamaki and K. Maida, Phys. Rev. D 62, 084041 (2000);

H. Yajima and T. Tamaki, Phys. Rev. D 63, 064007 (2001).

\bibitem{Kar05} S. Kar and S. Majumdar, Int. J. Mod. Phys. A 21, 6087 (2006).


\bibitem{Brigante} M. Brigante, H. Liu, R. C. Myers, S. Shenker and S.
Yaida, Phys. Rev. D 77, 126006 (2008);

Y. Kats and P. Petrov, JHEP 01, 044 (2009).

\bibitem{Kovtun} P. Kovtun, D. T. Son and A. O. Starinets, JHEP 10, 064
(2003).

\bibitem{CaiJHEP08} R-G. Cai and Y-W. Sun, JHEP 09, 115 (2008).

\bibitem{GeJHEP08} X-H. Ge, Y. Matsuo, F-W. Shu, S-J. Sin and T. Tsukioka,
JHEP 10, 009 (2008).

\bibitem{Jing10} J. Jing and S. Chen, Phys. Lett. B 686, 68 (2010).

\bibitem{Gregory09} R. Gregory, S. Kanno and J. Soda, JHEP 10, 010 (2009).

\bibitem{Pan10} Q. Y. Pan, B. Wang, E. Papantonopoulos, J. Oliveira and A.
Pavan, Phys. Rev. D 81, 106007 (2010).

\bibitem{Salim55} H. J. Mosquera Cuesta and J. M. Salim, Mon. Not. Roy.
Astron. Soc. 354, L55 (2004).

\bibitem{Salim25} H. J. Mosquera Cuesta and J. M. Salim, Ap. J. 608, 925
(2004).

\bibitem{BTZ1} M. Banados, C. Teitelboim and J. Zanelli, Phys. Rev. Lett.
69, 1849 (1992).

\bibitem{BTZ2} M. Banados, M. Henneaux, C. Teitelboim and J. Zanelli, Phys.
Rev. D 48, 1506 (1993).

\bibitem{BTZ3} S. Nojiri and S. D. Odintsov, Mod. Phys. Lett. A 13, 2695
(1998);

R. Emparan, G. T. Horowitz and R. C. Myers, JHEP 0001, 021 (2000);

S. Hemming, E. Keski-Vakkuri and P. Kraus, JHEP 0210, 006 (2002);

M. R. Setare, Class. Quantum Gravit. 21, 1453 (2004);

B. Sahoo and A. Sen, JHEP 0607, 008 (2006);

M. R. Setare, Eur. Phys. J. C 49, 865 (2007);

M. Cadoni and M. R. Setare, JHEP 0807, 131 (2008);

M. Park, Phys. Rev. D 77, 026011 (2008);

M. Park, Phys. Rev. D 77, 126012 (2008);

J. Parsons and S. F. Ross, JHEP 0904, 134 (2009);

M. R. Setare and M. Jamil, Phys. Lett. B 681, 469 (2009).

\bibitem{Carlip95} S. Carlip, Class. Quantum Gravit. 12, 2853 (1995).

\bibitem{Ashtekar02} A. Ashtekar, Adv. Theor. Math. Phys. 6, 507 (2002).

\bibitem{Sarkar06} T. Sarkar, G. Sengupta and B. Nath Tiwari, JHEP 0611, 015
(2006);

\bibitem{Witten98} E. Witten, Adv. Theor. Math. Phys. 2, 505 (1998).

\bibitem{Carlip05} S. Carlip, Class. Quantum Gravit. 22, R85 (2005).

\bibitem{Witten07} E. Witten, [arXiv:07063359].

\bibitem{Thermodynamics} S. P. Kim, S. K. Kim, K. S. Soh and J. H. Yee,
Phys. Rev. D 55, 2159 (1997);

G. W. Gibbons, M. J. Perry and C. N. Pope, Class. Quantum Gravit. 22, 1503
(2005);

J. E. Aman and N. Pidokrajt, Phys. Rev. D 73, 024017 (2006).

\bibitem{Saida99} H. Saida and J. Soda, Phys. Lett. B 471, 358 (2000).

\bibitem{Cadoni08} M. Cadoni, M. Melis and M. R. Setare, Class. Quantum
Gravit. 25, 195022 (2008).

\bibitem{Larranaga10} A. Larranaga, [arXiv:10023416].

\bibitem{Hyun97} S. Hyun, J. Korean Phys. Soc. 33, S532 (1998).

\bibitem{Sfetsos98} K. Sfetsos and K. Skenderis, Nucl. Phys. B 517, 179
(1998).

\bibitem{Canfora10} F. Canfora and A. Giacomini, [arXiv:10050091].

\bibitem{Claessens09} L. Claessens, [arXiv:09122245];

L. Claessens, [arXiv:09122267].

\bibitem{BTZlike} S. H. Hendi, Eur. Phys. J. C 71, 1551 (2011).

\bibitem{Liu26} J. T. Liu and P. Szepietowski, Phys. Rev. D 79, 084042
(2009);

Y. Kats, L. Motl and M. Padi, JHEP 0712, 068 (2007);

D. Anninos and G. Pastras, JHEP 0907, 030 (2009);

R. G. Cai, Z. Y. Nie and Y. W. Sun, Phys. Rev. D 78, 126007
(2008).

\bibitem{Gross87} D. J. Gross and J. H. Sloan, Nucl. Phys. B 291, 41 (1987).

\bibitem{Bergshoeff} E. A. Bergshoeff and M. de Roo, Nucl. Phys. B 328, 439
(1989).

\bibitem{Chemissany07} W. A. Chemissany, M. de Roo and S. Panda, JHEP 0708,
037 (2007).

\bibitem{Natsuume} M. Natsuume, Phys. Rev. D 50, 3949 (1994).

\bibitem{GHS} D. Garfinkle, G. T. Horowitz and A. Strominger, Phys. Rev. D
43, 3140 (1991) [Erratum ibid. 45, 3888 (1992)].

\bibitem{Ritz} A. Ritz and R. Delbourgo, Int. J. Mod. Phys. A 11, 253 (1996).

\bibitem{Heisenberg} W. Heisenberg and H. Euler, Z. Phys. 98, 714 (1936);
Translation by W. Korolevski and H. Kleinert, [physics/0605038]

\bibitem{Altshuler} B. L. Altshuler, Class. Quantum Gravit. 7, 189 (1990).

\bibitem{GibHaw} R. C. Myers, Phys. Rev. D 36, 392 (1987);

S. C. Davis, Phys. Rev. D 67, 024030 (2003).

\bibitem{BIBTZ} M. Cataldo and A. Garcia, [arXiv:hep-th/9903257];

R. Yamazaki and D. Ida, [arXiv:gr-qc/0105092];

Y. S. Myung, Y. W. Kim, and Y. J. Park, [arXiv:0804.0301].

\bibitem{Lambert} R. M. Corless, G. H. Gonnet, D. E. G. Hare, D. J. Jeffrey
and D. E. Knuth, Adv. Computational Math. 5, 329 (1996).

\bibitem{Gub} M. Cvetic and S. S. Gubser, JHEP. 04, 024 (1999);

M. M. Caldarelli, G. Cognola and D. Klemm, Class. Quantum Gravit.
17, 399 (2000).

\bibitem{Bekenstein} J. D. Bekenstein, Lett. Nuovo Cimento 4, 737 (1972);

J. D. Bekenstein, Phys. Rev. D 7, 2333 (1973);

S. W. Hawking and C. J. Hunter, Phys. Rev. D 59, 044025 (1999).

\bibitem{Hawking3} S. W. Hawking, C. J. Hunter and D. N. Page, Phys. Rev. D
59, 044033 (1999);

R. B. Mann, Phys. Rev. D 60, 104047 (1999);

R. B. Mann, Phys. Rev. D 61, 084013 (2000);

C. J. Hunter, Phys. Rev. D 59, 024009 (1999).
\end{thebibliography}
\end{document}